\documentclass[a4paper]{article}

\usepackage[utf8]{inputenc}

\usepackage{color}   
\usepackage[dvipsnames]{xcolor}
\usepackage{cite}
\usepackage{hyperref}
\hypersetup{
    colorlinks=true,     
    linktoc=all,         
    linkcolor=black,     
    citecolor=BlueViolet,
    urlcolor=BlueViolet  
}

\usepackage{tikz}
\usetikzlibrary{decorations.pathmorphing,arrows.meta,calc}
\usepackage{graphicx}
\graphicspath{ {./images/} } 

\usepackage{wrapfig}
\usepackage[rightcaption]{sidecap}

\usepackage{amsmath,amssymb,amsthm,amsfonts,amscd}
\usepackage{euscript}
\usepackage{wasysym} 
\usepackage{authblk}
\usepackage{lipsum}
\usepackage{geometry}
\usepackage{pdflscape}
\newgeometry{margin=2.55cm}
\usepackage{bm}

\theoremstyle{plain}
\newtheorem{theorem}{Theorem}[section]

\theoremstyle{definition}

\theoremstyle{remark}

\begin{document}

\title{\textbf{\large Cut-Out Wedges in $H_{3}$ and \\ the Borel-Resurgent Chern-Simons Matrix Integrals}}

\author[]{\normalsize Tuo Jia}
\affil[]{\normalsize Institute for Quantum Science and Engineering (IQSE)  \authorcr and Department of Physics and Astronomy, \authorcr Texas A\&M University, College Station, TX 77843, USA}

\author[]{\normalsize Zhaojie Xu}
\affil[]{\normalsize Department of Physics, \authorcr Shanghai University, Shanghai 200444, China}

\date{}

\maketitle

\begin{abstract}
In this paper, we present a systematic study of the Chern--Simons theory with gauge group \(\mathrm{SL}(2,\mathbb{R})\times\mathrm{SL}(2,\mathbb{R})\)  restricted to a wedge-identified manifold in the hyperbolic upper-half-space. The wedge geometry is created by imposing an angular cutoff in the \((x,y)\) plane and identifying two boundary lines, which introduces a single noncontractible loop in the manifold. By imposing the flat-connection condition of the Chern--Simons gauge fields, the path integral reduces to a finite-dimensional matrix integral in  \(\mathrm{SL}(2,\mathbb{R})\times\mathrm{SL}(2,\mathbb{R})\) . Although Chern-Simons theory is a topological theory, the resulting matrix integral remains nontrivial due to noncompact directions and boundary constraints.

The large-\(k\) expansion of the matrix integral is carried out by selecting a classical configuration in the space of holonomies and expanding around it in inverse powers of \(k\). The resulting coefficients of the asymptotic series exhibit factorial growth,  enabling us to apply the Borel resummation techniques.  Summation over these subleading sectors removes potential ambiguities in the Borel integral and clarifies the emergence of a resurgent transseries structure. In the Borel-resurgent analysis, we show that, despite the apparent simplicity of the reduced action, the wedge geometry yields a rich interplay of perturbative and non-perturbative phenomena. This work presents an explicit example of how a finite-dimensional matrix integral in its expansions is physically meaningful through Borel resummation.
\end{abstract}

\tableofcontents

\section{Introduction}
\label{sec:1-intro}

Chern--Simons theory \cite{Witten:1988hf, Witten:2010cx, Terashima:2011qi, Gang:2017hbs} is a three-dimensional topological quantum field theory \cite{Witten:1988ze} defined on a three-manifold \(M\) and governed by a gauge connection \(A\) taking values in the Lie algebra of a chosen gauge group \(G\). One writes the classical action in the form
\begin{equation}
S_{\mathrm{CS}}[A]
=
\frac{k}{4\pi}
\int_M
\mathrm{Tr}
\Bigl(
A \wedge \mathrm{d}A
\;+\;
\tfrac{2}{3}\,A \wedge A \wedge A
\Bigr),
\end{equation}
where \(k\in \mathbb{Z}\) is the Chern--Simons level (or coupling), and \(\mathrm{Tr}\) denotes an invariant quadratic form on the Lie algebra of \(G\). In the simplest cases, one can take \(G\) to be \(\mathrm{SU}(N)\), \(\mathrm{U}(1)\), or \(\mathrm{SO}(N)\); however, it can be any compact or noncompact group with a well-defined invariant bilinear form.

Chern--Simons theory has broad applications in mathematics and physics. Mathematically, it underlies the study of knot invariants \cite{Witten:1988hc, Garoufalidis:2021osl} (such as the Jones polynomial) and three-manifold invariants. While in physics, it  shows up in descriptions of fractional statistics, topological orders in condensed matter systems \cite{Dunne:1998qy, PhysRevB.81.195303}.  In three-dimensional gravity with negative cosmological constant, one can use the first-order (Chern--Simons) formulation \cite{Witten:1988hc, Witten:2007kt}. Therefore, it is interesting to study the formulations and the structures of Chern-Simons theory.

The primary goal of this paper is to formulate and analyze the path integral of Chern--Simons theory with gauge group \(\mathrm{SL}(2,\mathbb{R})\times\mathrm{SL}(2,\mathbb{R})\) \cite{Witten:1988hc, Witten:2007kt, Yin:2007gv, Maloney:2007ud, PhysRevLett.115.161304, PhysRevD.93.064014} in $H_{3}$, focusing on a specific cut-out wedge manifold in the hyperbolic upper-half-space model of Euclidean \(\mathrm{AdS}_{3}\). Although the local degrees of freedom vanish, the global structures, holonomies around non-contractible loops, and subtleties at boundaries or asymptotic regions lead to intriguing quantum phenomena. In our analysis, the theory is reduced to a finite-dimensional matrix integral over the group \(\mathrm{SL}(2,\mathbb{R})\times\mathrm{SL}(2,\mathbb{R})\), one can study asymptotic expansions in the Chern--Simons level \(k\) and observe a resurgent structure upon performing Borel resummation    \cite{Gukov:2016njj, Garoufalidis:2020xec, Garoufalidis:2021osl, Duan:2022ryd, Benjamin:2023uib, Benjamin:2024cvv}.

Euclidean AdS\(_{3}\) or $H_{3}$ can be represented by AdS\(_{3}\) under a Wick rotation, equipped with the upper-half-space coordinates \((x,y,z)\) for \(z>0\). If a loop is introduced by restricting angular variables and making identifications in the \((x,y)\) plane, that identification can yield a nontrivial fundamental group \(\pi_{1}(M)\). Then, the resulting manifold is topologically distinct from the simply connected \(H_{3}\). In this paper, we perform precisely such a restriction, namely cutting out a wedge by choosing \(0 \le \theta \le \Delta\theta\) in polar coordinates and identifying \(\theta=0\) with \(\theta=\Delta\theta\). This gluing procedure leaves us with a manifold that has one noncontractible loop and hence a single holonomy in each \(\mathrm{SL}(2,\mathbb{R})\) factor. Imposing flatness on the gauge connections reduces the infinite-dimensional path integral to a finite-dimensional matrix integral over those monodromies.

The boundary data required to make the matrix integral finite or regulated is encoded in the constraints on the traces of these monodromies, for instance \(\mathrm{Tr}(M) = 2\,\cosh(\alpha)\) in the hyperbolic sector. In our case, we fix \(\mathrm{Tr}(M)=2\,\cosh(\alpha)\) and \(\mathrm{Tr}(\bar{M})=2\,\cosh(\bar{\alpha})\). The resulting integral, while finite-dimensional, can still exhibit divergences or indefinite directions in \(\mathrm{SL}(2,\mathbb{R})\) unless the measure is properly constrained. Nonetheless, at large \(k\), a formal expansion in powers of \(1/k\) can be carried out by picking a classical solution \((M_{0},\bar{M}_{0})\) that stationarizes the reduced action. In this analysis, such expansions diverge factorially in the number of loops or corrections, enabling us to undertake a Borel summation and obtain the transseries that include all subleading exponentials.

The phenomenon of factorially divergent expansions is well known in quantum field theory. One frequently encounters it in perturbation expansions for gauge theories or in topological quantum field theories where instanton-like saddles appear. The wedge geometry introduces the boundary or global effects by providing a nontrivial monodromy, which then leads to nontrivial expansions around different saddle points in \(\mathrm{SL}(2,\mathbb{R})\times \mathrm{SL}(2,\mathbb{R})\).

One of the most stimulating outcomes of this perspective is that we obtain a concrete, finite-dimensional testing ground for the resurgent viewpoint in 
 topological quantum field theory and even in quantum gravity. At a superficial glance, the reduced action \(\Phi_{\mathrm{wedge}}(M,\bar{M})\) can appear deceptively simple: for instance, in a hyperbolic holonomy scenario, \(\mathrm{Tr}(M)=2\,\cosh(\alpha)\) can reduce the exponent of the integrand to something like \(\alpha-\bar{\alpha}\) up to factors. Nonetheless, precisely because \(\mathrm{SL}(2,\mathbb{R})\) is noncompact and boundary constraints must be imposed, the matrix integral can exhibit subtle divergences and an intricate factorial growth in the loop expansions. Thus, far from being trivial, this matrix integral for the wedge geometry is a rich example of how topological theories unify perturbative and non-perturbative effects.

A further implication is that correlators or boundary excitations can also be included, if desired, though we primarily focus on the pure partition function (or vacuum amplitude) perspective in this  analysis. Exploring correlation functions with the insertion of Wilson lines or boundary WZW fields \cite{PhysRevD.100.126009} could expand the scope of these methods.

The remainder of this paper is organized as follows. Section 2 reviews Chern--Simons theory in Euclidean  AdS\(_{3}\), explaining how flatness translates to a moduli of homomorphisms from \(\pi_{1}(M)\) into \(\mathrm{SL}(2,\mathbb{R})\times \mathrm{SL}(2,\mathbb{R})\). Section 3 defines the wedge geometry in hyperbolic space, sets up the gluing identification \(\theta=0\sim\theta=\Delta\theta\), and shows how the path integral is reduced to a matrix integral over \((M,\bar{M})\). Section 4 performs the large-\(k\) expansions, identifies factorial growth of loop coefficients, and applies Borel summation, discussing the appearance of subleading saddles. Section 5 concludes with a summary of how this wedge-geometric model provides a concrete example of resurgent behavior in a topological field theory setting, along with potential avenues for future exploration such as boundary WZW matter, black-hole analogs, or correlation function insertions.

\section{Reviewing Chern--Simons Theory}
\label{sec:2-review}

In this section, we provide a more detailed review of Chern--Simons theory.  The gauge group being considered of the Chern--Simons theory is \(\mathrm{SL}(2,\mathbb{R})\times \mathrm{SL}(2,\mathbb{R})\). Denoting the gauge fields by \(A\) and \(\bar{A}\), we write the combined Chern--Simons action in the form
\begin{equation}
S_{\mathrm{CS}}[A,\bar{A}]
=
\frac{k}{4\pi}
\int_{M}
\mathrm{Tr}\Bigl(A\wedge \mathrm{d}A+\tfrac{2}{3} A\wedge A\wedge A\Bigr)
-
\frac{k}{4\pi}
\int_{M}
\mathrm{Tr}\Bigl(\bar{A}\wedge \mathrm{d}\bar{A}+\tfrac{2}{3}\,\bar{A}\wedge \bar{A}\wedge \bar{A}\Bigr).
\end{equation}

The manifold \(M\) in our considerations will be a three-dimensional space of Euclidean signature, realized as a cut-out region of hyperbolic space. One varies the total action with respect to \(A\),
\begin{equation}
F(A) = \mathrm{d}A + A\wedge A = 0,
\end{equation}
and similarly varying with respect to \(\bar{A}\) imposes
\begin{equation}
F(\bar{A}) = 0.
\end{equation}
These are the flatness conditions on the two gauge fields. As a topological theory, the local degrees of freedom vanish, but there remain important global effects captured by the fundamental group of \(M\) and the corresponding monodromies of \(A,\bar{A}\).

To see how the path integral can be reduced, note that in Chern--Simons theory, the gauge field one-form \(A\) is integrated over all configurations modulo gauge equivalences. If the manifold is simply connected and has no boundary, one can write \(A = g^{-1} \mathrm{d}g\). However, once there is a nontrivial fundamental group or boundary conditions/constraints, the space of flat connections can be parametrized by homomorphisms
\begin{equation}
\rho: \ \pi_{1}(M)\ \longrightarrow \ \mathrm{SL}(2,\mathbb{R}),
\quad
\bar{\rho}:\ \pi_{1}(M)\ \longrightarrow\ \mathrm{SL}(2,\mathbb{R}),
\end{equation}
modded out by conjugations in each factor. This shows that local degrees of freedom vanish, and only the global holonomies contribute. Once the path integral is restricted to the flat sector, one obtains a finite-dimensional integral or sum (depending on discrete or continuous factors) over possible group elements representing the holonomies of \(\rho\) and \(\bar{\rho}\).

The boundary can further enforce constraints such as fixing certain cycles to have specific lengths or specifying the boundary metric. In Euclidean AdS\(_3\), if one cuts out a portion and identifies edges, then the path integral picks up a factor from holonomies around the newly introduced loops. More precisely, if \(\pi_{1}(M)\) is isomorphic to \(\mathbb{Z}\) or some other group, each generator corresponds to an \(\mathrm{SL}(2,\mathbb{R})\) element in the first factor and another in the second. Denote these elements by \(M,\bar{M}\). Imposing boundary or wedge conditions will restrict \(\mathrm{Tr}(M),\mathrm{Tr}(\bar{M})\) to lie in a certain range (for example, hyperbolic with \(\mathrm{Tr}(M)>2\)).

The total action is
\begin{equation}
S_{\mathrm{CS}}[A,\bar{A}]
=
\frac{k}{4\pi}
\Bigl[\,
\int \mathrm{Tr}(A\wedge \mathrm{d}A + \tfrac{2}{3}A^{3})
\Bigr]
-
\frac{k}{4\pi}
\Bigl[\,
\int \mathrm{Tr}(\bar{A}\wedge \mathrm{d}\bar{A} + \tfrac{2}{3}\bar{A}^{3})
\Bigr].
\end{equation}
Since each of \(A,\bar{A}\) is the flat connection in classical solutions, the main content often arises from boundary terms, or from the discrete holonomy data. In three dimensions, these discrete data reflect the entire geometry: a single loop can produce a single pair of monodromies \((M,\bar{M})\). The path integral in this setting can be schematically written as
\begin{equation}
Z(M) 
=
\int_{\{\text{flat }A,\bar{A}\}}
\exp\bigl\{
i\,S_{\mathrm{CS}}[A,\bar{A}]
\bigr\},
\end{equation}
and therefore be reduced to an integral over group elements \((M,\bar{M})\in \mathrm{SL}(2,\mathbb{R})\times\mathrm{SL}(2,\mathbb{R})\) . The explicit measure emerges from quotienting by gauge equivalences, and boundary conditions further impose constraints, such as requiring \(\mathrm{Tr}(M)=2\cosh(\alpha)\). Because of the noncompactness of \(\mathrm{SL}(2,\mathbb{R})\), one obtains indefinite integrals unless these constraints or partial gauge fixings are introduced.

In the following sections, we place this formulation into our specific wedge geometry in \(H_{3}\). That geometry has a single nontrivial loop due to angular identification in the plane, ensuring that \((M,\bar{M})\) parametrizes the flat connections up to gauge transformations. We then show how, at large Chern--Simons level \(k\), one can systematically expand the resulting finite-dimensional integrals in inverse powers of \(k\), observe the factorial growth of the coefficients, and perform a Borel resummation to interpret the full expansion.

\section{Cut-Out Wedge in \texorpdfstring{$H_3$}{H\_3} and the Reduction to a Matrix Integral}
\label{sec:3-wedge}

In this section, we introduce the cut-out wedge geometry in the upper-half-space model of \(H_3\) and show how the path integral of the Chern--Simons theory is reduced to a finite-dimensional matrix integral over the monodromies. We first describe the wedge construction, then identify the single noncontractible loop, and finally impose boundary constraints on \(\mathrm{Tr}(M)\) and \(\mathrm{Tr}(\bar{M})\).

\subsection{Defining the Wedge Geometry in \texorpdfstring{$H_{3}$}{H\_3}}

We start with the upper-half-space model of the three-dimensional hyperbolic space, where coordinates \((x,y,z)\) satisfy \(z>0\) and the Euclidean AdS\(_{3}\) (or hyperbolic) metric is
\begin{equation}
\mathrm{d}s^{2}
=
\frac{\mathrm{d}x^{2}+\mathrm{d}y^{2}+\mathrm{d}z^{2}}{z^{2}}.
\end{equation}
One can use planar polar coordinates \((r,\theta)\) in the \((x,y)\) plane so that
\begin{equation}
x
=
r\cos(\theta),
\quad
y
=
r\sin(\theta),
\quad
z>0.
\end{equation}

In a completely uncut hyperbolic space, \(\theta\) would range from \(0\) to \(2\pi\), but we now restrict the domain to an interval \(0 \le \theta \le \Delta\theta\) with \(0 \le r <\infty\) and \(z>0\). The important step here is to glue/identify the boundary line at \(\theta=0\) with the boundary line at \(\theta=\Delta\theta\). This identification means that a path in the \((x,y)\) plane that increments \(\theta\) from \(0\) to \(\Delta\theta\) returns to the same physical location after crossing the boundary, thus producing a noncontractible loop in the resulting manifold.

Simply put, one ``cuts out'' all of hyperbolic space except for the region in that restricted angular sector, then identifies the two radial boundary lines. The manifold that remains after this restriction and identification is referred to as the wedge manifold, and it has the property that its fundamental group \(\pi_{1}\) is isomorphic to \(\mathbb{Z}\), generated by the loop that goes around in \(\theta\mapsto\theta+\Delta\theta\). This is the topological feature that will later allow us to reduce the Chern--Simons path integral to a finite-dimensional integral over the resulting holonomy.

To visualize this, in the figure below we show a simplified diagram of the wedge manifold as seen in a 3D perspective with the coordinates \((r,\theta)\) in the plane and the vertical axis for \(z>0\). One can see that restricting \(0\le \theta\le \Delta\theta\) corresponds to cutting out all angles beyond \(\theta=\Delta\theta\), and then gluing that cut line to \(\theta=0\). The manifold that remains is topologically different from the full simply connected \(H_{3}\). This geometric construction is the essential ingredient in producing a single loop that the flat connections can wind around.

\bigskip

\begin{figure}[htbp]
\centering
\begin{tikzpicture}[scale=1.1]

\coordinate (O) at (0,0);
\draw[->, thick] (0,0) -- (3.5,0) node[right] {$r$};
\draw[->, thick] (0,0) -- (0,3) node[above] {$z>0$};

\begin{scope}[shift={(0,0)}]
\draw[thick, blue] (0,0) -- ++(50:3) node[pos=0.65,above,xshift=2pt] {$\theta=\Delta\theta$};
\draw[thick, blue] (0,0) -- ++(0:3) node[pos=0.65,below] {$\theta=0$};
\draw[blue] (1.2,0) arc[start angle=0, end angle=50, radius=1.2];
\node[blue] at ($(0.9,0)+(25:1.15)$) {$\Delta\theta$};

\draw[dashed] (3,0) -- (3,1.8);
\draw[dashed] ($(50:3)$) -- ($(50:3)+(0,1.8)$);

\node at (3.6,1.2) {Wedge region in the plane};
\end{scope}

\end{tikzpicture}
\caption{A schematic 2D cross-sectional view showing how the wedge is defined in the \((r,\theta)\) plane, with $z>0$ up the vertical axis. One restricts $0\le \theta\le \Delta\theta$, then identifies the boundary $\theta=0$ with $\theta=\Delta\theta$, producing a single noncontractible loop.}
\label{fig:WedgeInH3}
\end{figure}
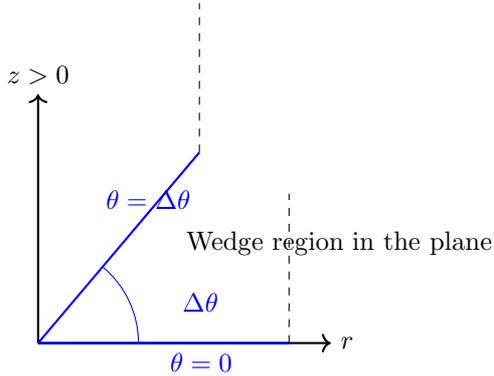

Once the wedge manifold is formed by this identification, one can embed it in the full $(x,y,z)$ upper-half-space. Locally the geometry remains hyperbolic, since no curvature is introduced, but globally we now have a loop on which a flat gauge connection can wind. This loop is exactly what generates a nontrivial holonomy once the Chern--Simons fields are integrated out, leading to the matrix integral valued in  $\mathrm{SL}(2,\mathbb{R})\times \mathrm{SL}(2,\mathbb{R})$.

\subsection{Noncontractible Loop and Holonomies}

Once the wedge manifold is formed by restricting $0 \le \theta \le \Delta\theta$ in the plane and identifying $\theta=0$ with $\theta=\Delta\theta$, one obtains a single noncontractible loop in the manifold, generated by a path $\gamma$ that increments $\theta$ from $0$ to $\Delta\theta$ while holding $r,z$ fixed in a small neighborhood. In more explicit terms, one may parametrize this loop as
\begin{equation}
\gamma:\ 
\theta \mapsto
\theta + \Delta\theta  \quad\text{(mod identification)}
\end{equation}
within the region of $H_3$ that remains. This loop $\gamma$ is noncontractible, and the manifold’s fundamental group $\pi_1(M_{\mathrm{wedge}})$ is isomorphic to $\mathbb{Z}$, generated by this loop.

For the single loop $\gamma$ in the wedge, the holonomy is
\begin{equation}
M
=
\rho(\gamma)
\in
\mathrm{SL}(2,\mathbb{R}).
\end{equation}
Similarly, in the second factor, we have
\begin{equation}
\bar{M}
=
\bar{\rho}(\gamma)
\in
\mathrm{SL}(2,\mathbb{R}).
\end{equation}
The path integral over flat connections thus reduces to an integral over the pair $(M,\bar{M})\in  \mathrm{SL}(2,\mathbb{R})\times\mathrm{SL}(2,\mathbb{R})$, subject to boundary conditions.

\medskip

\noindent
\textbf{Relation between \boldmath$\Delta\theta$ and the Hyperbolic Parameters \boldmath$\alpha,\bar{\alpha}$.}
In the hyperbolic (AdS) scenario, one typically fixes the trace of each monodromy in a real, greater-than-2 range. This yields
\begin{equation}
\mathrm{Tr}(M)
=
2\,\cosh(\alpha),
\quad
\mathrm{Tr}(\bar{M})
=
2\,\cosh(\bar{\alpha}),
\end{equation}
with real $\alpha,\bar{\alpha}$. If we interpreted $\Delta\theta$ as a literal \emph{real} angle of rotation in the plane, that would correspond to an \emph{elliptic} element in $\mathrm{SL}(2,\mathbb{R})$ with $\mathrm{Tr}(M)=2\cos(\Delta\theta)$. However, here we have a \emph{hyperbolic} identification, so in the simplest setting, one can formally treat
\begin{equation}
\Delta\theta
=
i\,\alpha
\quad
\text{(``rotation'' by an imaginary angle, i.e.\ a real hyperbolic boost).}
\end{equation}

\medskip

\noindent
\textbf{Significance in the Path Integral.}
Imposing $\mathrm{Tr}(M)=2\cosh(\alpha)$ and $\mathrm{Tr}(\bar{M})=2\cosh(\bar{\alpha})$ is equivalent to stating the wedge loop $\gamma$ is realized by a hyperbolic holonomy of size $\alpha$ (and $\bar{\alpha}$ for the second factor). Such boundary data, once implemented in the path integral, localizes the measure in $\mathrm{SL}(2,\mathbb{R}) \times \mathrm{SL}(2,\mathbb{R})$, removing indefinite directions and leading to a partially regulated matrix integral.  The wedge loop is the  noncontractible cycle, and $\alpha,\bar{\alpha}$ are the real parameters fixing each monodromy’s hyperbolic length. In the large-\(k\) expansions we will analyze later, $\alpha,\bar{\alpha}$ remain real, and the wedge geometry’s $\Delta\theta$ is effectively an \emph{imaginary} extension if one tries to interpret it as a rotation.

\subsection{Matrix Integral and Boundary Constraints on \texorpdfstring{$\mathrm{Tr}(M)$}{Tr(M)}}

Once the wedge identification is introduced in \(H_3\) and the gauge connections \(A,\bar{A}\) are required to be flat, the full Chern--Simons path integral is reduced to an integral over the monodromies \((M,\bar{M})\in  \mathrm{SL}(2,\mathbb{R})\times\mathrm{SL}(2,\mathbb{R})\). We denote this path integral by 
\begin{equation}
Z_{\mathrm{wedge}}(k)
=
\int_{\mathrm{SL}(2,\mathbb{R}) \times \mathrm{SL}(2,\mathbb{R})}
\mathrm{d}M\,\mathrm{d}\bar{M}
\;
\exp\Bigl\{
i\,k\,\Phi_{\mathrm{wedge}}(M,\bar{M})
\Bigr\},
\end{equation}
where \(\Phi_{\mathrm{wedge}}(M,\bar{M})\) is the ``reduced action'' obtained by evaluating the Chern--Simons functional on flat connections with monodromies \(M,\bar{M}\).

In principle, the group measure \(\mathrm{d}M\,\mathrm{d}\bar{M}\) is derived from quotienting out all gauge redundancies and local degrees of freedom, leaving only the global variables characterizing the flat connection. However, \(\mathrm{SL}(2,\mathbb{R})\) is noncompact, so if no further conditions are imposed, the integral can formally diverge or become indefinite. Physically, in Euclidean AdS\(_3\), one typically enforces boundary conditions that fix certain geometric parameters corresponding to geodesic lengths or black-hole horizon data. Concretely, these boundary conditions often specify the trace of each monodromy.

To fix the length-like parameter \(\alpha\) for the monodromy \(M\in\mathrm{SL}(2,\mathbb{R})\), we set
\begin{equation}
\mathrm{Tr}(M)
=
2\,\cosh(\alpha).
\end{equation}
Similarly for \(\bar{M}\), the parameter \(\bar{\alpha}\) yields 
\begin{equation}
\mathrm{Tr}(\bar{M})
=
2\,\cosh(\bar{\alpha}).
\end{equation}
Thus, if one wants to incorporate those constraints directly into the integral, one can introduce delta functions enforcing these trace conditions:
\begin{equation}
\delta\bigl(\mathrm{Tr}(M)-2\cosh(\alpha)\bigr)
\quad\text{and}\quad
\delta\bigl(\mathrm{Tr}(\bar{M})-2\cosh(\bar{\alpha})\bigr).
\end{equation}
The matrix integral thus becomes

\begin{align}
Z_{\mathrm{wedge}}(k) &= \int_{ \mathrm{SL}(2,\mathbb{R})\times\mathrm{SL}(2,\mathbb{R})}
\mathrm{d}M\,\mathrm{d}\bar{M}
\,
\delta\bigl(\mathrm{Tr}(M)-2\cosh(\alpha)\bigr)
\,
\delta\bigl(\mathrm{Tr}(\bar{M})-2\cosh(\bar{\alpha})\bigr)
\; \notag \\
     &\times \exp\Bigl\{
i\,k\,\Phi_{\mathrm{wedge}}(M,\bar{M})
\Bigr\}. \label{eq:wedgeaction}
\end{align}
Reducing the theory to the saddle points (which will be explained in more detail in Section 4), one obtains  
\begin{equation}
\Phi_{\mathrm{wedge}}(M_0,\bar{M}_0)
=
\alpha-\bar{\alpha}
\end{equation}

In many treatments, these constraints ensure that the wedge loop is assigned a precise hyperbolic ``boost'' amplitude \(\alpha\) (and \(\bar{\alpha}\) for the second factor), correlating with how the wedge boundary is realized in the Euclidean AdS geometry. For instance, in a black-hole analogy, one might interpret \(\alpha\) as a real function of the horizon circumference or mass parameter, but in simpler wedge contexts, it simply becomes the real length-like parameter controlling the monodromy.

One may also partially gauge-fix the conjugation freedoms in \(\mathrm{SL}(2,\mathbb{R})\). The group measure \(\mathrm{d}M\) can be ill-defined if the entire group is integrated over without quotienting out redundancies or restricting to a submanifold defined by the trace condition. Imposing 
\[
\mathrm{Tr}(M)
=
2\,\cosh(\alpha)
\quad
(\text{and similarly for $\bar{M}$})
\]
removes one dimension from \(\mathrm{SL}(2,\mathbb{R})\). One is then typically left with an integration over the coset describing how \(M\) is conjugated, although additional boundary or horizon-like constraints may reduce that further. The goal is to tame the noncompactness sufficiently so that the integral, while still indefinite in certain directions, can be tackled via large-\(k\) expansions.

In section 4, we will expand 
\begin{equation}
Z_{\mathrm{wedge}}(k)
=
\int
\mathrm{d}M\,\mathrm{d}\bar{M}\,
\exp\Bigl\{
i\,k\,\Phi_{\mathrm{wedge}}(M,\bar{M})
\Bigr\}
\end{equation}
around the saddle points \((M_{0},\bar{M}_{0})\) that extremize \(\Phi_{\mathrm{wedge}}(M,\bar{M})\).  Small fluctuations around $M_0$ then produce integrals over $\delta\mathcal{X}\in\mathfrak{sl}(2,\mathbb{R})$, with the partition function expansions exhibiting factorial growth in loop order. This is the reason we see large-\(k\) divergences that require Borel resummation, resulting in the resurgent structure once subleading monodromy sectors are included in the total sum.

\subsection{The Reduction of Path Integral and Interpretation}

In this subsection, we  explain how the original three-dimensional Chern--Simons path integral, which is a priori an integral over an infinite-dimensional space of gauge fields, reduces to a finite-dimensional matrix integral once we enforce the flatness condition in the wedge manifold. We also discuss why this reduced integral admits a direct geometric interpretation that clarifies the role of boundary constraints.

In the usual presentation of Chern--Simons theory, one typically integrates over gauge fields \(A\) and \(\bar{A}\) in \(\mathrm{SL}(2,\mathbb{R})\times \mathrm{SL}(2,\mathbb{R})\) on a manifold \(M\). In our case, \(M\) is the wedge manifold cut out from the upper-half-space \(H_3\) by restricting \(0\le \theta\le \Delta\theta\) and identifying the boundary lines \(\theta=0\) and \(\theta=\Delta\theta\). If these gauge fields are not constrained, one faces a complicated functional integral. However, the classical Chern--Simons equations of motion, which enforce \(F(A)=0\) and \(F(\bar{A})=0\), make all local curvature vanish. In a simply connected manifold, this would imply each connection is pure gauge, but here the manifold has a single noncontractible loop,  which corresponds to the gluing identification. This loop allows for nontrivial monodromies \((M,\bar{M})\in \mathrm{SL}(2,\mathbb{R})\times\mathrm{SL}(2,\mathbb{R}) \), ensuring that the manifold is not topologically trivial.

Given that a flat \(\mathrm{SL}(2,\mathbb{R})\) connection is fully determined by the group element it assigns to each generator of \(\pi_1(M)\), the presence of exactly one noncontractible loop reduces the entire path integral to a finite-dimensional integral over \((M,\bar{M})\). Symbolically, we shift from an integral \(\int [\mathcal{D}A\,\mathcal{D}\bar{A}]\) to \(\int \mathrm{d}M\,\mathrm{d}\bar{M}\). In this wedge manifold with boundary or some other constrained conditions, one typically inserts further constraints on \(\mathrm{Tr}(M)\) and \(\mathrm{Tr}(\bar{M})\). These constraints remove directions associated with indefinite volume in \(\mathrm{SL}(2,\mathbb{R})\). The resulting finite-dimensional integral may still have divergences if additional gauge quotients are not taken, but it is still drastically simpler than any infinite-dimensional path integral.

Because the local degrees of freedom are absent in three dimensions, there is no separate ``bulk field'' integral left after imposing the flat connection condition. Instead, the entire dynamical content is encoded in the boundary data and the global holonomies.  If one defines a boundary condition that fixes \(\mathrm{Tr}(M)=2\cosh(\alpha)\) and \(\mathrm{Tr}(\bar{M})=2\cosh(\bar{\alpha})\), then one captures the length-like parameters \(\alpha,\bar{\alpha}\) in the geometry. Physically, these 
 may be interpreted as geodesic lengths or horizon circumferences, depending on the context. The classical solution that stationarizes the reduced action \(\Phi_{\mathrm{wedge}}(M,\bar{M})\) can then be identified in \(\mathrm{SL}(2,\mathbb{R})\times \mathrm{SL}(2,\mathbb{R})\), which leads to a well-defined saddle-point approach.

\section{Asymptotic Expansions, Borel Summation, and Resurgent Structure}
\label{sec:4-resurgence}

In this section, we analyze the large-\(k\) expansions of the finite-dimensional matrix integral arising from the wedge geometry, and then we demonstrate the factorially growing feature of the resulting coefficients in the series, and apply Borel resummation techniques to show the resurgent structure \cite{Dorigoni:2014hea, Dunne:2015eaa, bridge, Costin:2023kla}. The main points are that the reduced action \(\Phi_{\mathrm{wedge}}(M,\bar{M})\) admits classical solutions \((M_0,\bar{M}_0)\) around which the integral can be expanded in inverse powers of \(k\),  the coefficients of that expansion grow like \(n!\), and the Borel transform of the series has singularities tied to subleading saddles.

\subsection{Large-\texorpdfstring{$k$}{k} Expansion Around a Classical Saddle}

The matrix integral arising from the wedge geometry is
\begin{equation}
Z_{\mathrm{wedge}}(k)
=
\int_{\mathrm{SL}(2,\mathbb{R}) \times \mathrm{SL}(2,\mathbb{R}) }
\mathrm{d}M\,\mathrm{d}\bar{M}
\,
\exp\Bigl\{
i\,k\,\Phi_{\mathrm{wedge}}(M,\bar{M})
\Bigr\},
\end{equation}
with the constraints \(\mathrm{Tr}(M)=2\cosh(\alpha)\) and \(\mathrm{Tr}(\bar{M})=2\cosh(\bar{\alpha})\). We extremize the reduced action
\begin{equation}
\frac{\partial \Phi_{\mathrm{wedge}}}{\partial M}\Bigl|_{(M_0,\bar{M}_0)}=0
\quad
\text{and}
\quad
\frac{\partial \Phi_{\mathrm{wedge}}}{\partial \bar{M}}\Bigl|_{(M_0,\bar{M}_0)}=0.
\end{equation}
The solution \((M_0,\bar{M}_0)\) satisfies $\mathrm{Tr}(M_0)=2\cosh(\alpha)$, $\mathrm{Tr}(\bar{M}_0)=2\cosh(\bar{\alpha})$, and we obtain the simple form
\begin{equation}
\Phi_{\mathrm{wedge}}(M_0,\bar{M}_0)
=
\alpha-\bar{\alpha}
\end{equation}
up to an overall sign and normalization conventions. In what follows, 
 we denote this classical value by
\begin{equation}
S_{0}
=
\Phi_{\mathrm{wedge}}(M_0,\bar{M}_0).
\end{equation}

To expand about this saddle, we write
\[
M
=
M_{0}\,\exp\bigl(\delta \bm{\xi}\bigr),
\quad
\bar{M}
=
\bar{M}_{0}\,\exp\bigl(\delta \bar{\bm{\xi}}\bigr),
\]
where $\delta \bm{\xi},\ \delta \bar{\bm{\xi}}\in \mathfrak{sl}(2,\mathbb{R})$ are infinitesimal deformations.  One inserts these into $\Phi_{\mathrm{wedge}}(M,\bar{M})$ and performs a Taylor expansion about $(M_0,\bar{M}_0)$. Let $\delta \bm{\xi}^a$ be a basis of coordinates for the space of fluctuations in $M$, and similarly $\delta \bar{\bm{\xi}}^{\bar{a}}$ for $\bar{M}$. Then we have
\begin{align}
\Phi_{\mathrm{wedge}}\bigl(M_{0} \exp(\delta \bm{\xi}),\;\bar{M}_{0} \exp(\delta \bar{\bm{\xi}})\bigr) &= \Phi_{\mathrm{wedge}}(M_0,\bar{M}_0)
+  \notag \\
     &= \sum_{a}\frac{\partial \Phi_{\mathrm{wedge}}}{\partial \bm{\xi}^a}\Bigl|_{(M_0,\bar{M}_0)}\delta \bm{\xi}^a
+      \notag \\
&  \sum_{\bar{a}}\frac{\partial \Phi_{\mathrm{wedge}}}{\partial \bar{\bm{\xi}}^{\bar{a}}}\Bigl|_{(M_0,\bar{M}_0)}\delta \bar{\bm{\xi}}^{\bar{a}}
+  \notag \\
&\text{higher orders}. \label{eq:expand}
\end{align}
By the stationary condition, the linear terms vanish at $(M_0,\bar{M}_0)$. The quadratic order involves integrals of the form
\begin{equation}
\frac12
\sum_{a,b}
\frac{\partial^{2} \Phi_{\mathrm{wedge}}}{\partial \bm{\xi}^a \partial \bm{\xi}^b}\Bigl|_{(M_0,\bar{M}_0)}
\;\delta \bm{\xi}^a\,\delta \bm{\xi}^b
+
\text{(cross terms with $\delta\bar{\bm{\xi}}$ and purely $\delta\bar{\bm{\xi}}$ terms)}.
\end{equation}
This second derivative is the Hessian of $\Phi_{\mathrm{wedge}}$ at the saddle. Higher orders in $\delta \bm{\xi},\delta \bar{\bm{\xi}}$ continue in a polynomial manner, leading to a sum over loop diagrams.

Thus, the exponent in the matrix integral becomes
\begin{equation}
i\,k
\Bigl[
S_0
+
Q\bigl(\delta \bm{\xi},\delta \bar{\bm{\xi}}\bigr)
+
\ldots
\Bigr]
\end{equation}
where $Q(\delta \bm{\xi},\delta \bar{\bm{\xi}})$ stands for the quadratic and higher-order expansions.

Denoting the measure by $\mathrm{d}(\delta \bm{\xi})\,\mathrm{d}(\delta \bar{\bm{\xi}})$ up to Jacobian factors from the group parametrization, one obtains
\begin{equation}
Z_{\mathrm{wedge}}(k)
=
\exp\{\,i\,k\,S_0\}
\int
\mathrm{d}(\delta \bm{\xi})\,\mathrm{d}(\delta \bar{\bm{\xi}})
\,
\exp\Bigl\{
i\,k\,Q(\delta \bm{\xi},\delta \bar{\bm{\xi}})+
i\,k\,(\text{cubic and higher terms})
\Bigr\}.
\end{equation}
We perturbatively expand the integral to obtain a power series
\begin{equation}
Z_{\mathrm{wedge}}(k)
=
\exp\bigl\{ i\,k\,S_0\bigr\}
\sum_{r=0}^\infty
a_r\,
\frac{1}{k^r}.
\end{equation}
Factorial growth arises from the combinatorial enumeration of ways to pick higher-order terms in $\delta \bm{\xi},\delta \bar{\bm{\xi}}$. 

Combining all the results, the wedge matrix integral is written as
\begin{equation}
Z_{\mathrm{wedge}}(k)
=
\exp\Bigl\{
i\,k\,(\alpha-\bar{\alpha})
\Bigr\}
\sum_{r=0}^\infty
a_r\,
\frac{1}{k^r}.
\end{equation}
The expansions in $\delta \bm{\xi}$ and $\delta \bar{\bm{\xi}}$ fix the coefficients $a_r$ precisely by integrating out all polynomial interactions in the exponent. This means $a_r$ is an integral over $(m+n)$-th powers of the fluctuations at loop order $r$. Symbolically, if $r$ counts how many powers of $\frac{1}{k}$ one picks up, each additional vertex in the expansion of $Q(\delta \bm{\xi},\delta \bar{\bm{\xi}})$ contributes combinatorial factors that yield $a_r\sim c^r\,r!$ for large $r$. As a result, the large-$k$ series diverges factorially, enabling us to use the Borel resummation to interpret the asymptotic expansions.

\subsection{Factorial Growth and the Borel Transform}

Here, we further illustrate the factorial growth of the coefficients $a_{n}$.  First, we notice that each order in \(1/k\) arises from integrating polynomial or rational expressions in the small fluctuations \(\delta\bm{\xi}, \delta\bar{\bm{\xi}}\in \mathfrak{sl}(2,\mathbb{R})\) over a Gaussian-like measure from the quadratic approximation of the exponent, plus subleading interaction terms. Concretely, if one writes the expanded action near the saddle as
\begin{equation}
i\,k
\Bigl(
S_0
+
Q_2(\delta\bm{\xi},\delta\bar{\bm{\xi}})
+
Q_3(\delta\bm{\xi},\delta\bar{\bm{\xi}})
+
\cdots
\Bigr),
\end{equation}
the subscript on \(Q_m\) denotes terms of $m$-th order in the small fluctuations. The integral measure over \(\delta\bm{\xi},\delta\bar{\bm{\xi}}\) produces loop diagrams in a matrix-model sense. At each loop order \(n\), we expand the exponential of $Q_3+Q_4+\dots$ to $n$-th power and compute correlations of the fluctuations under the Gaussian given by $Q_2$. Symbolically, we obtain integrals such as
\begin{equation}
\int
(\delta \bm{\xi})^m
(\delta \bar{\bm{\xi}})^p
\,
\exp\Bigl\{
i\,k\,Q_2(\delta \bm{\xi},\delta \bar{\bm{\xi}})
\Bigr\},
\end{equation}
multiplied by combinatorial coefficients which are from the series expansion of 
\(\exp\{i\,k\,Q_3+\cdots\}\). The main point here is that at the $n$-th loop order in $1/k$, one must pick $n$ powers from these interaction vertices in $Q_3, Q_4, \dots$, leading to factorial combinatorial growth as $n$ increases.

Thus, a statement is given, 
\begin{equation}
a_n
\sim
(\mathrm{const.})^n\,
n!,
\end{equation}
since each additional order in $n$ opens up a combinatorial factor that scales roughly like $n!$\footnote{For the ``$\mathrm{const.}$'' in $(\mathrm{const.})^{n}$,  at one loop,  it is $\sim$ (the one-loop determinant) $\times$ (phase factors) $\times$ (other possible measure factors). When we go to higher loops, since the integral is still being taken over a finite-dimensional measure, and we still have the invertibility of the Hessain matrix.  Each additional loop order is just another polynomial factor in a finite set of integration variables.  Moreover, the wedge geometry is well-defined, each saddle is properly gauge-fixed, an infinite gauge volume will not appear and no zero modes remain.  Therefore, the ``$\mathrm{const.}$'' is still well-bounded at higher loops. Hence, this resulting matrix model is \emph{Borel-resummable.}}.  Here, despite the matrix integral being finite-dimensional, the expansions in $\delta \bm{\xi}$, $\delta \bar{\bm{\xi}}$ for higher-order interaction terms replicate the same combinatorial escalation, reminiscing the results from expansions in other quantum field theories or matrix models. One concludes that the series
\begin{equation}
\sum_{n=0}^\infty
a_n\,\frac{1}{k^n}
\end{equation}
diverges factorially.

The standard resolution is to define a Borel transform that reorganizes the divergent sum into a possibly convergent integral. To be precise, we let
\begin{equation}
f(k)
=
\sum_{n=0}^\infty
a_n\,
\frac{1}{k^n}.
\end{equation}
Its \emph{Borel transform} is given by
\begin{equation}
\widehat{\mathcal{B}}[f](t)
=
\sum_{n=0}^\infty
\frac{a_n}{n!}\; t^n.
\end{equation}
Because $a_n\sim (c)^n\,n!$, the ratio $a_n/n!\sim c^n$ ensures that $\widehat{\mathcal{B}}[f](t)$ has at least some finite radius of convergence around $t=0$. One then attempts to recover $f(k)$ from $\widehat{\mathcal{B}}[f](t)$ by an inverse Laplace transform in $t$, known as the \emph{Borel sum}:
\begin{equation}
f_{\mathrm{Borel}}(k)
=
\int_0^\infty
\mathrm{d}t\,
\exp(-k\,t)
\,
\widehat{\mathcal{B}}[f](t).
\end{equation}
If $\widehat{\mathcal{B}}[f](t)$ is free of singularities for $0\le t<\infty$ on the real axis, this integral uniquely defines the sum of the divergent series $\sum_{n=0}^\infty a_n\,k^{-n}$. However, in some cases, one often finds singularities at $t^*=i(S_\alpha-S_0)$ or real axis branch points if $\Delta S=S_\alpha-S_0$ is purely imaginary, corresponding to subleading classical saddles in the wedge geometry. In that scenario, one must define a lateral Borel sum and deal with an imaginary discontinuity that is then canceled by the expansions of subleading saddles in a full transseries approach.

We have shown that the factorial growth is a direct reflection of the combinatorial escalation of terms in the zero-dimensional expansions of the wedge matrix integral, and the Borel transform 
\begin{equation}
\widehat{\mathcal{B}}[f](t)
=
\sum_{n=0}^\infty
\frac{a_n}{n!}
\,t^n
\end{equation}
is the standard method to reorganize the asymptotic series. Together with subleading exponentials, we will have a resurgent unification of perturbative and non-perturbative effects in the wedge-Euclidean AdS$_3$ model.

\subsection{Subleading Saddles and Resurgent Bridging}

We consider the perturbation series around a principal saddle, with classical action denoted by \(S_{0}\). When expanded in powers of \(1/k\), the Borel transform of this principal-saddle series may have singularities lying on or intersecting the real axis in the Borel plane. These real-axis singularities make  the integral
\begin{equation}
\int_{0}^{\infty}
e^{-\,k\,t}
\,
\widehat{\mathcal{B}}[f](t)\,\mathrm{d}t
\end{equation}
 be multi-valued unless one stipulates whether the integration contour passes slightly above or slightly below the branch or pole. This multi-valuedness manifests as an imaginary discontinuity in the partial Borel sums of the principal saddle,  written as 
\begin{equation}
Z_{\mathrm{main}}^{(+)}(k)
\;-\;
Z_{\mathrm{main}}^{(-)}(k),
\end{equation}
where \(\bigl(Z_{\mathrm{main}}^{(\pm)}(k)\bigr)\) denotes the partial Borel sums taken above or below the real axis. The difference is a purely imaginary quantity that shows an intrinsic ambiguity in the principal-saddle series when viewed in isolation.

A main insight of the transseries formalism is that if there exist further saddle points, each with an action
\begin{equation}
S_{\alpha}
=
S_{0} + \Delta S_{\alpha},
\end{equation}
then each such saddle contributes an exponential term
\begin{equation}
\exp\bigl\{\,i\,k\,S_{\alpha}\bigr\}
\sum_{m=0}^{\infty}
b_{\alpha,m}\,
\frac{1}{k^{m}}
\end{equation}
to the full path integral. In resurgent analysis, one finds that the Borel-plane singularities in the principal saddle's partial sums arise precisely at locations tied to \(\Delta S_{\alpha}\). The discontinuity of the principal-saddle series at that singularity encodes the leading amplitude of the subleading saddle. In the alien derivative language \cite{Dorigoni:2014hea, Bellon:2016med}, the large-order divergence of one saddle ``knows about'' the leading exponential factor from another.

This interplay is referred to as \emph{resurgent bridging}. It is stated the imaginary jump in the partial Borel sums of the principal saddle is precisely canceled once the subleading exponential is added. A succinct way of expressing this is
\begin{equation}
Z_{\mathrm{main}}^{(+)}(k)
-
Z_{\mathrm{main}}^{(-)}(k)
=
\exp\bigl\{\,i\,k\,S_{\alpha}\bigr\}
\Bigl[
b_{\alpha,0} + O\!\bigl(\tfrac{1}{k}\bigr)
\Bigr],
\end{equation}
where \(b_{\alpha,0}\) is the leading amplitude determined by the local Gaussian integral (and any gauge-fixing determinants) around \((M_{\alpha},\bar{M}_{\alpha})\). Including subleading powers in \(1/k\) on both sides systematically refines this equality. Thus, the real-axis ambiguity in the principal saddle’s partial sum is removed by the subleading exponential factor, ensuring that after all subleading saddles have been accounted for in the total transseries, no real-axis discontinuity remains.

Hence, \emph{resurgent bridging} refers to the exact cancellation of the principal-saddle partial-sum ambiguity by exponential contributions from subleading saddles. This resolves the multi-valuedness of the principal-saddle expansions, leading to an unambiguous real result once one sums the entire transseries (across all saddles, each with its loop expansions). The bridging mechanism does not require that the subleading expansions match the principal-saddle series at every term. Instead, it guarantees that the difference between lateral Borel sums in the principal sector is offset by a carefully dictated amount from the leading term of the subleading exponentials, so that the final amplitude no longer depends on how one avoids the real-axis singularity in the Borel transform.

\section{Conclusion and Future Perspectives}

The wedge-geometric Chern--Simons theory described in this work highlights a rich interplay between global monodromies, boundary conditions, and resurgent asymptotic expansions. By restricting the three-dimensional manifold to a wedge and identifying its boundary lines, one is left with a single or finite number of noncontractible loops, ensuring that the path integral, subject to the flatness condition, collapses to a finite-dimensional integral over the monodromies in \(\mathrm{SL}(2,\mathbb{R}) \times \mathrm{SL}(2,\mathbb{R})\). Although the resulting theory is topological and has no local propagating degrees of freedom, its perturbative expansions in powers of \(1/k\) around each saddle give us factorially divergent series.

The high-order growth of these series shows that simple truncations of perturbation theory are inadequate on their own. Resurgent analysis provides the framework to address this divergence and interpret the full transseries solution. In particular, the presence of multiple saddles, each contributing an exponential factor, explains why singularities arise in the Borel plane of the principal saddle’s perturbation series. These singularities are then canceled/bridged by the leading contributions of subleading exponentials, ensuring that no real-axis ambiguity persists once all saddle expansions are included. This mechanism, known as resurgent bridging, makes the total amplitude unambiguous and free of imaginary discontinuities that would otherwise plague partial Borel sums.

The wedge model thereby exemplifies the general resurgent scenario in a simplified, finite-dimensional setting. Its holonomy/monodromy-based path integral reveals the structure seen in more complicated quantum field theories, where multiple nontrivial saddles and topological configurations lead to similar factorial divergences and Borel-plane singularities. This wedge-based construction thus clarifies how transseries, encompassing all saddles, unify the various expansions into a single coherent resurgent framework.

Going forward, one natural direction is to investigate correlation functions or insertion operators in the wedge geometry, where boundary conditions like WZW fields  \cite{PhysRevD.100.126009}  or Wilson lines might introduce additional structure in the finite-dimensional integral. Another interesting and promising  avenue is to refine the transseries analysis by incorporating subleading loops more explicitly for each saddle. Such refinements will enable us to see the full resurgent patterns emerge in greater detail, addressing how each real-axis singularity in the principal saddle’s Borel transform is successively neutralized at higher orders by subleading exponentials. A deeper comparison with analogous lens-space Chern--Simons partitions and with lower-dimensional models might reveal universal features of factorial divergence and resurgent cancellations. Overall, the wedge-geometric approach offers a clear, localized setting in which to study these phenomena, providing a bridge between topological quantum field theory and resurgent theory.

\subsection*{Acknowledgments}
TJ is supported by the Herman F. Heep and Minnie Belle Heep Texas A\&M University Endowed Fund held/administered by the Texas A\&M Foundation.

\appendix

\section{Appendix: Technical Details and Supplementary Computations}

In the appendix, we provide a number of detailed derivations and clarifications that can provide background for some of the arguments presented in the main text.

\medskip

\noindent
\textbf{A. Parametrizations in \(\mathrm{SL}(2,\mathbb{R})\) for Hyperbolic Elements}

\medskip

\noindent
Let M $\in \mathrm{SL}(2,\mathbb{R})$
be a hyperbolic element so that \(\mathrm{Tr}(M) = 2\cosh(\alpha)\) for a real \(\alpha\). A canonical normal form is
\begin{equation}
M
=
P
\begin{pmatrix}
e^{\alpha} & 0 \\
0 & e^{-\alpha}
\end{pmatrix}
P^{-1},
\end{equation}
where \(P\in \mathrm{SL}(2,\mathbb{R})\). One can introduce a delta function \(\delta(\mathrm{Tr}(M)-2\cosh(\alpha))\) to implement the constraint, thereby localizing to a two-dimensional submanifold in the three-dimensional group. The remaining gauge freedom is a quotient by conjugations that identify \(M \sim Q\,M\,Q^{-1}\). Once such freedoms are accounted for, the integral over \(\mathrm{SL}(2,\mathbb{R})\) effectively reduces to a line or circle in the space of eigenvalues plus a coset space for the matrix \(P\).

To see how this works in the wedge integral, one writes 
\begin{equation}
\int_{\mathrm{SL}(2,\mathbb{R})} 
\mathrm{d}\mu(M)\;
\delta\bigl(\mathrm{Tr}(M)-2\cosh(\alpha)\bigr),
\end{equation}
possibly multiplied by a gauge quotient factor. Therefore, the noncompactness of \(\mathrm{SL}(2,\mathbb{R})\) is partially tamed, and a real parameter \(\alpha\) can be integrated over if the boundary data do not pin it to a fixed value. A second factor \(\bar{M}\) in the gauge group is handled similarly, with \(\mathrm{Tr}(\bar{M})=2\cosh(\bar{\alpha})\).

\medskip

\noindent
\textbf{B. The Combinatorial Structure of the Loop Expansions Near Classical Saddles}

\medskip

\noindent
Consider a saddle \((M_0,\bar{M}_0)\). We have 
\begin{equation}
M = M_0 \exp(\delta \bm{\xi}),
\quad
\bar{M} = \bar{M}_0 \exp(\delta \bar{\bm{\xi}}).
\end{equation}
This introduces fluctuations in a small neighborhood of the classical point. The action near this point expands as
\begin{equation}
\Phi_{\mathrm{wedge}}(M,\bar{M})
=
\Phi_{\mathrm{wedge}}(M_0,\bar{M}_0)
+
\Phi^{(2)}(\delta \bm{\xi}, \delta \bar{\bm{\xi}})
+
\Phi^{(3)}(\delta \bm{\xi}, \delta \bar{\bm{\xi}})
+
\cdots,
\end{equation}
where \(\Phi^{(2)}\) is quadratic (akin to a Gaussian), while \(\Phi^{(3)}, \Phi^{(4)}, \dots\) are higher-order interaction vertices in the zero-dimensional sense. The wedge integral thus takes the form
\begin{equation}
\exp\bigl\{i\,k\,S_0\bigr\}
\int
\exp\bigl\{
i\,k\,\bigl[\Phi^{(2)} + \Phi^{(3)} + \cdots\bigr]
\bigr\}
\mathrm{d}(\delta \bm{\xi})\,\mathrm{d}(\delta \bar{\bm{\xi}}).
\end{equation}
At order \(n\) in \(1/k\), one selects \(n\) powers of the higher-order terms from the expansion of \(\exp\{i\,k(\Phi^{(3)}+\cdots)\}\), each of which is integrated against the Gaussian from \(\Phi^{(2)}\). The number of such terms grows combinatorially in \(n\). When one sums over all possible ways to pick the powers and all possible contractions of \(\delta \bm{\xi},\delta \bar{\bm{\xi}}\), the coefficient at order \(1/k^n\)  gives an \(n!\)-type factor.

Thus, the partial sums of the loop expansion yield factorial growth in their high-order coefficients, validating the statement that each $a_n \sim (c)^n\,n!$ for large $n$, leading to a divergent asymptotic series.

\medskip

\noindent
\textbf{C. Enumerating Subleading Solutions and Their Action Differences}

\medskip

\noindent
A subleading saddle \((M_\alpha,\bar{M}_\alpha)\) might differ from the principal one by shifting \(\alpha\mapsto \alpha + 2\pi i n\) (similarly for \(\bar{\alpha}\)), or by changing the sign or orientation in the wedge identification. If one normalizes the wedge action so that \(\mathrm{Tr}(M)=2\cosh(\alpha)\) leads to \(\Omega(M)=\alpha\) for the boundary functional, then each shift 
\begin{equation}
\alpha' = \alpha + 2\pi i n
\end{equation}
can in principle produce a new classical monodromy with an action 
\begin{equation}
S_\alpha'
=
\Omega(M_\alpha') - \Omega(\bar{M}_\alpha')
\quad
\text{or a variant.}
\end{equation}
The difference \(\Delta S = S_\alpha' - S_0\) becomes purely imaginary if \(n\) is an integer, typically placing the resulting branch point along the real axis of the Borel plane at $t= \Re(i\Delta S)$. One then sees how such a discrete set of subleading solutions can generate a tower of possible singularities. In practice, not all are necessarily relevant if certain boundary data break those conditions, but the principle stands that the relevant subleading saddles each appear in the transseries and remove the partial Borel-sum ambiguities.

\medskip

\noindent
\textbf{D. On the Existence and Enumerability of Subleading Saddles}

\medskip

\noindent
For the \(\mathrm{SL}(2,\mathbb{R})\times \mathrm{SL}(2,\mathbb{R})\) Chern--Simons theory on a wedge manifold, 
it is standard to encounter a primary (or principal) saddle corresponding to monodromies 
\begin{equation}
M_0,\;\bar{M}_0
\in
\mathrm{SL}(2,\mathbb{R}),
\end{equation}
with boundary constraints of the form \(\mathrm{Tr}(M_0)=2\cosh(\alpha)\), \(\mathrm{Tr}(\bar{M}_0)=2\cosh(\bar{\alpha})\). This Theorem establishes that there can be a discrete (finite or countably infinite) family of \emph{subleading saddles} arising from integer ``boosts'' or complex shifts of \(\alpha,\bar{\alpha}\), each yielding a legitimate classical solution and therefore contributing an exponential factor in the total path integral.

\medskip

\begin{theorem}[Existence and Enumerability of Subleading Saddles]
\label{thm:SubleadingSaddles}
Let \(M_0,\bar{M}_0\in\mathrm{SL}(2,\mathbb{R})\) be the holonomies in the wedge geometry of Euclidean AdS\(_3\), forming a classical solution that stationarizes the reduced action \(\Phi_{\mathrm{wedge}}(M,\bar{M})\). Assume \(M_0,\bar{M}_0\) are  \emph{hyperbolic} so that 
\begin{equation}
\mathrm{Tr}(M_0) = 2\cosh(\alpha),
\quad
\mathrm{Tr}(\bar{M}_0) = 2\cosh(\bar{\alpha}),
\end{equation}
with real \(\alpha,\bar{\alpha}\). Suppose the global boundary conditions (and any wedge identifications) allow one to define the following family of group elements for each integer \(n\in\mathbb{Z}\), 
\begin{equation}
\alpha_n = \alpha + 2\pi i n,
\quad
\bar{\alpha}_n = \bar{\alpha},
\end{equation}
yielding candidate holonomies \(M_n,\bar{M}_n\in\mathrm{SL}(2,\mathbb{R})\) satisfying the same stationarity condition up to boundary identifications. Then:

\begin{enumerate}
\item For each integer \(n\), the pair \((M_n,\bar{M}_n)\) either forms a legitimate classical saddle of \(\Phi_{\mathrm{wedge}}(M,\bar{M})\) or fails to meet the wedge boundary constraints. In the former case, we call it a \emph{subleading} (or ``shifted'') saddle.  
\item The collection of all such \((M_n,\bar{M}_n)\) that remain solutions is at most \emph{countably infinite} in \(n\).  
\item If the wedge boundary conditions indeed permit the integer shift in the hyperbolic exponent as a genuine solution, then each valid \(n\) corresponds to a distinct monodromy, giving a distinct classical action 
\begin{equation}
S_n
=
\Phi_{\mathrm{wedge}}(M_n,\bar{M}_n),
\end{equation}
and therefore a subleading exponential factor \(\exp\{i\,k\,S_n\}\).  
\end{enumerate}
\end{theorem}

\medskip

\noindent
\textbf{Proof.}     Since \(M_0\) is hyperbolic with \(\mathrm{Tr}(M_0)=2\cosh(\alpha)\), there exists a decomposition 
\begin{equation}
M_0 = P\,\mathrm{diag}(\,e^\alpha,\,e^{-\alpha}\,)\,P^{-1}
\end{equation}
for some \(P\in\mathrm{SL}(2,\mathbb{R})\).  The wedge boundary identification or stationary condition enforces that \(M_0\) solve certain equations derived from \(\delta \Phi_{\mathrm{wedge}}=0\). Now consider 
\(
\alpha_n
=
\alpha + 2\pi i n
\)
for integer \(n\). Define 
\begin{equation}
M_n
=
P\,
\mathrm{diag}\bigl(e^{\alpha_n},\,e^{-\alpha_n}\bigr)
\,
P^{-1}.
\end{equation}
Then \(\mathrm{Tr}(M_n)=2\cosh(\alpha_n)\). Note that \(\cosh(\alpha + 2\pi i n)=\cosh(\alpha)\) if the ``\(\cosh\)'' function is extended to complex arguments in the usual way, which means \(\mathrm{Tr}(M_n)=2\cosh(\alpha)\) in a purely algebraic sense. 

Notice that not every \(M_n\) is a solution to the wedge boundary conditions. For instance, if the boundary constraints also fix geodesic lengths or angles in a purely real sense, one might lose solutions for large imaginary shifts. However, the set of candidates 
\[
\{\,(M_n,\bar{M}_n)\mid n\in \mathbb{Z}\}
\]
is at most countably infinite (since \(n\) is an integer). One either discards those that do not meet the wedge boundary constraints or keeps those that do.

To put in a more concrete note, if for some integer \(n\), the pair \((M_n,\bar{M}_n)\) also solves the stationarity equations 
\begin{equation}
\frac{\partial \Phi_{\mathrm{wedge}}}{\partial M}(M_n,\bar{M}_n)=0,
\quad
\frac{\partial \Phi_{\mathrm{wedge}}}{\partial \bar{M}}(M_n,\bar{M}_n)=0,
\end{equation}
then \((M_n,\bar{M}_n)\) is a legitimate classical saddle. Otherwise, it is ruled out. Thus, we will form a subset 
\begin{equation}
\mathcal{S}
=
\bigl\{ n\in\mathbb{Z}\,\mid (M_n,\bar{M}_n)\text{ is a valid wedge solution}\bigr\}.
\end{equation}
Since \(\mathbb{Z}\) is countably infinite, \(\mathcal{S}\subseteq \mathbb{Z}\) is also \emph{at most countably infinite}.  

If $(M_n,\bar{M}_n)$ and $(M_{n'},\bar{M}_{n'})$ differ by $n'\neq n$, one will obtain distinct monodromies (except in degenerate cases where conjugation or equivalences might identify them). Thus, they yield distinct classical actions 
\begin{equation}
S_n
=
\Phi_{\mathrm{wedge}}(M_n,\bar{M}_n).
\end{equation}

In conclusion, we have shown that the set of possible subleading saddles generated by integer shifts in $\alpha$ or $(\alpha,\bar{\alpha})$ is at most countably infinite. If the wedge boundary conditions indeed allow each $\alpha_n,\bar{\alpha}_n$ to remain a legitimate solution, then each $n\in\mathcal{S}$ yields a subleading saddle with classical action $S_n$.

\qedsymbol

\medskip

\noindent
\textbf{Remark.}  
The crucial step lies in showing that the shift $\alpha\mapsto\alpha+2\pi i\,n$ (or variants) does not violate any wedge boundary constraints and that the stationarity condition remains satisfied. If, in a given geometry, such conditions fail beyond $n=\pm 1$ or fail entirely for $n\neq0$, then no additional saddles appear. But if, for instance, the wedge geometry does allow these imaginary shifts in the exponent to maintain the same boundary conditions, one obtains countably many subleading saddles. Either way, the important property is the \emph{discreteness} of $n\in\mathbb{Z}$, ensuring at most a countable family.

\medskip

\noindent
\textbf{F. Additional Remarks on Residual Gauge Fixings and Measures}

\medskip

\noindent
Lastly, one wonders how the measure in \(\mathrm{SL}(2,\mathbb{R})\times \mathrm{SL}(2,\mathbb{R})\) fully factors out the gauge transformations. Usually one writes 
\begin{equation}
\int 
[\mathcal{D}A\,\mathcal{D}\bar{A}]
\quad\longmapsto\quad
\int_{\mathrm{flat/gauge}}
\mathrm{d}M\;\mathrm{d}\bar{M},
\end{equation}
where ``flat/gauge'' indicates one has used \(F(A)=0, F(\bar{A})=0\) plus gauge transformations to reduce all local field degrees of freedom to a finite set of global monodromies. In a typical wedge scenario, boundary constraints further specify \(\mathrm{Tr}(M)\) or \(\mathrm{Tr}(\bar{M})\), removing noncompact integration directions. While the exact measure can be really complicated, the final expansions often do not hinge on its explicit form; one  can obtain it up to a constant or finite factor that does not affect the factorial divergence in the loop expansions or the locations of Borel-plane singularities.

\end{document}